\documentclass[aps,pre,
superscriptaddress,groupedaddress,showpacs,onecolumn]{revtex4}
\usepackage{graphicx,epsf}
\usepackage{hyperref}
\usepackage{xcolor}
\usepackage{psfrag}
\usepackage{psfrag}
\usepackage[english]{babel}
\usepackage[title]{appendix}
\begin{document}
\title{Molecular conformations of dumbbell-shaped polymers in good solvent}
\author{Khristine Haydukivska}
\affiliation
{Institute of Physics, University of Silesia, 41-500 Chorz\'ow, Poland}
\affiliation{Institute for Condensed
Matter Physics of the National Academy of Sciences of Ukraine,\\
79011 Lviv, Ukraine}

\author{V. Blavatska}
\affiliation{Institute for Condensed
Matter Physics of the National Academy of Sciences of Ukraine,\\
79011 Lviv, Ukraine}
\affiliation{Dioscuri Centre for Physics and Chemistry of Bacteria,
Institute of Physical Chemistry, Polish Academy of Sciences, 01-224 Warsaw, Poland}
\author{Jaros{\l}aw Paturej}
\email[]{E-mail:  jaroslaw.paturej@us.edu.pl}
\affiliation
{Institute of Physics, University of Silesia, 41-500 Chorz\'ow, Poland}
\affiliation{Leibniz-Institut f\"ur Polymerforschung Dresden e.V., 01069 Dresden,
Germany}

\begin{abstract}
We study conformational properties of diluted dumbbell polymers which consist of two rings that are attached to both ends of a linear spacer segment by using  
analytical methods of field theory and bead-spring coarse-grained molecular dynamics simulations. 
We investigate the influence of the relative length of the spacer segment to the length of side rings on the shape and the relative size of dumbbells as compared to linear polymers of equal mass.
We find that dumbbells with short spacers are much more compact than linear polymers.  
Oppositely, we observe that the influence of side rings on the size of   dumbbells becomes negligible with increasing length of a spacer. Consequently  dumbbell molecules with long spacers become comparable in size with corresponding linear chains. Our analytical theory predicts quantitative cross-over between short- and long-spacer behavior and is confirmed by numerical simulations.      
\end{abstract}
\date{\today}
\maketitle

\section{Introduction}
\begin{figure}
    \centering
    \includegraphics[width=73mm]{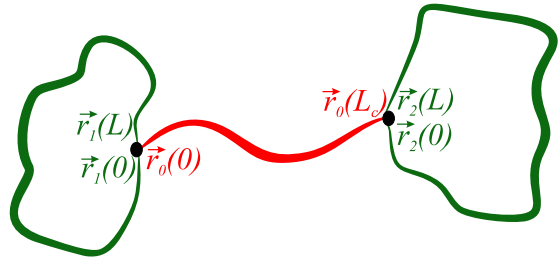}
    \caption{Schematic representation of dumb-bell polymer with side rings. }
    \label{DB_sheme}
\end{figure}    

In recent decades a lot of attention is directed to the research of polymers with complex architectures since they are characterised by unique physical properties. Particularly interesting are molecular architectures  with no free ends with the simplest example being ring polymers, that have been thoroughly investigated\cite{Roovers88,McLeish2002,Duplantier}. Ring polymers are observed in nature in bacteria\cite{Fiers62} and in some higher eukaryotes\cite{Zhou03} which contain a circular DNA. Synthetic ring polymers are known to have a smaller size \cite{Zimm49,Douglas84} in solutions as well as no rubbery plateau in melts \cite{Doi15}.  Mixtures of closed and opened chains demonstrate a number of interesting behaviours in melts\cite{Halverson2012}. However such mixtures are somewhat limited and it was suggested that tad-pole polymers with chains and rings chemically connected have a better threading topological constraints and provides a wider range of possible dynamic behaviour\cite{Rosa20}. 

On another hand a wide range of research into the reology of pom-pom polymers was conducted in recent years\cite{Ruymbeke,McLeish98,Graham01}. With side arms influencing the reology of the melt\cite{Ruymbeke}. 

In a recent study \cite{Doi21} a system of dumbbell polymers (see fig. \ref{DB_sheme}) was investigated experimentally in both melt and solution. The presence of rings on both ends of the chain when they are are threaded by the other dumbbell molecules is leading to the heavily constrained motion and thus a slow dynamic. On the other hand those same molecules in dilute solution have a similar behavior to that of the simple chains, in particular the similar values of viscosity where observed. 

Also polymer melts and dense solutions form the core of material applications, dilute solutions have their small roles as well, for example as a viscosity modifiers\cite{Shell,Vlassopoulos14,Roovers88}, as complex polymers are usually characterised by a lower intrinsic viscosity than their linear counterparts of the same mass. This is usually connected with the decrease  in effective polymer size, with relation between intrinsic viscosity and effective size being described by the Flory-Fox equation \cite{Berry2007,Kok1981}. Another important goal that can be achieved by studying dilute solutions is that it allows to study the properties of individual molecules as in dilute solution the interactions between different macromolecules are neglectfully small\cite{Burchard}.

The decrease in  size for brunched structures is measured by shrinking factor that is usually considered as a ratio between  intrinsic viscosities of branched and linear macromolecules of the same molecular mass. In analytical descriptions a related ratio between the gyration radiuses is considered since the pioneering work of Zimm and Stockmayer\cite{Zimm49}:
\begin{equation}
g_c = \frac{\langle R^2_g\rangle}{\langle R^2_g\rangle_{\rm linear}}    \label{gc}
\end{equation}
Another example of the universal characteristics is asphericity ($\langle A_d \rangle$)~\cite{Aronowitz}. The quantity $A_d$ provides description of the shape of polymer configuration and distinguishes between spherical ($\langle A_d \rangle=0$) and  rod-like configurations ($\langle A_d \rangle=1$). The asphericity is defined as:
\begin{equation}
\langle A_d\rangle=\frac{1}{d(d-1)}\left\langle\frac{{\rm Tr} \hat{\bf S}^2 }{({\rm Tr} {\bf S})^2} \right\rangle\label{Ad}
\end{equation}
where $ {\bf S}$ in the gyration tensor, $\hat{ {\bf S}}= {\bf S}-\overline{\mu}{\bf I}$ with $\overline{\mu}$ being an average eigenvalue and ${\bf I}$ a unity matrix. 
The quantities $g_c$ and $A_d$ are example of the so-called universal characteristics, that depends only on global properties of the system. 
In the case of polymers in dilute solutions such characteristic depends on the space dimension, quality of solvent, polymer branching and architecture but does not depend on the details of the monomer chemistry.  

With a number of strategies  for synthesis of complex polymers with rings developed \cite{jan2017} in recent decades including dumbbell\cite{Tezuka2013,oike200}, sometimes called manacle-shape polymers\cite{Doi21}, a theoretical consideration of universal characteristics may be of some academic interest. 

In this work we consider the dumbbell polymers using both analytical and numerical approaches. We star with the short description of the methods used in this research in Section \ref{Methods} that is followed by discussion of the results in Section \ref{Reslts}. We close this paper with concluding remarks in Section \ref{Con}. 

\section{Models and Methods} 
\label{Methods}
\subsection{Analytical model}
An analytical description is conducted using field-theoretical continuous chain model \cite{Edwards}. In this model polymer chain is represented by a trajectory of length $L$ parameterized by the radius vector $\vec{r}(s)$ where $s$ varies from $0$ to $L$. The Hamiltonian of the model is given as:
\begin{eqnarray}
&&H = \frac{1}{2}\sum_{i=1}^{F}\,\int_0^{L_i} ds\,\left(\frac{d\vec{r_i}(s)}{ds}\right)^2\nonumber\\
&&+\frac{u}{2}\sum_{i,j=1}^{F}\int_0^{L_i}ds'\int_0^{L_j} ds''\,\delta(\vec{r_i}(s')-\vec{r_j}(s'')).\label{H}
\end{eqnarray}
In the above equation $F$ denotes functionality, i.e. a number of trajectories in the branched polymer architecture (for a dumbbell polymer it is $F=3$) and $u$ is a coupling constant that describes the strength of the excluded volume interactions. The polymer topology  is introduced in the partition function of the system by fixing end(s) of trajectories:
\begin{eqnarray}
&&Z^{{\rm DB}}_{L_c,L}=\frac{1}{Z_0^{{\rm DB}}}\int\,D\vec{r}(s)\times\delta(\vec{r_1}(0)-\vec{r_0}(0))\delta(\vec{r_2}(0)-\vec{r_0}(L_c))\times\nonumber\\
&&\delta(\vec{r_1}(0)-\vec{r_1}(L))\delta(\vec{r_2}(0)-\vec{r_2}(L))\,{\rm e}^{-H},
\label{ZZ}
\end{eqnarray}
where  $\delta(\vec{r_1}(0)-\vec{r_0}(0))$ and $\delta(\vec{r_2}(0)-\vec{r_0}(L_c))$ describe the connectivity of ring trajectories parameterized by vectors $\vec{r}_1$ and $\vec{r}_2$ to the ends of the linear trajectory $\vec{r}_0$ and the remaining two  $\delta$-functions impose geometrical constraints to trajectories subjected them to form closed rings (see Fig.~\ref{DB_sheme}). 
Note that in our study we consider dumb-bell topologies with trajectories of different length. The linear backbone trajectory  has the length $L_c$ whereas the trajectory of side rings has the length $L$. We also point out that since both $L$ and $L_c$ in our analytical model are considered to be infinitely long the difference in length is introduced by considering the ratio $\lim_{L,L_c\rightarrow\infty} L_c/L=l$.
We also note that in the continuous chain model the averaging over  possible configurations that is performed in the partition function includes all types of knot conformations of the rings. Consequently these conformations are not distinguishable in our model.  However since the rings are considered to be infinitely long  \cite{desCloiseaux} the knots are localized and do not influence neither scaling exponents nor the critical amplitudes  \cite{Douglas10,Kantor,Orlandini}.

In the continuous chain model the contribution from the excluded volume interactions is considered to be much smaller as compared to the Gaussian elasticity term.  As a consequence all the observables are calculated as a perturbation series over the coupling constant $u_0$ \cite{desCloiseaux}. In general the partition function has the following form:
\begin{equation}
Z(L,L_c)=Z_0\left(1-u_0 Z_1(l,d)+\ldots\right)    \label{Z}
\end{equation}
where $u_0=u(2\pi)^{-\frac{d}{2}}L^{2-\frac{d}{2}}$ is a dimensionless coupling constant.  In the the case of a dumbbell polymer $Z_0=(2\pi L)^{-d}$. The second coefficient $Z_1(l,d)$ in the above equation represents the contribution from the excluded volume interactions and is calculated using the diagrammatic technique. The final expression for the coefficient $Z_1(l,d)$ as well as the details of calculations are provided in the Appendix (cf.~Eq.~(\ref{Z1appendix}))  All the observables can be written in a form of Eq.~(\ref{Z}). In particular the radius of gyration is given as:
\begin{equation}
\langle R^2_g\rangle=\langle R^2_g\rangle_0\left(1-u_0 R_1(l,d)+\ldots\right)\label{R2g}    
\end{equation}
where $\langle R^2_g\rangle_0$ denotes the contribution from the Gaussian conformation and the term $R_1$ is the first-order approximation of steric interactions. The  explicit formula for term $R_1$ is provided in the Eq.~(\ref{R1explicit}) and the sketch of calculation of this term is given in the Appendix. 

In the continuous chain model all observables are depend on the coupling constant $u_0$ that diverges as $L\rightarrow\infty$. In order to calculate finite physical value of the observable a renormalization has to be introduced such that $u^*_0 \rightarrow u^*_{R}$ as the $L\rightarrow\infty$. The direct polymer renormalization approach for the model given by the Hamiltonian (cf.~Eq.~\ref{H}) leads to the following fixed points~\cite{desCloiseaux}:
\begin{eqnarray}
&& {\rm {Gaussian}}: u^*_{\rm{R}}=0,\qquad {\text at}\qquad d\geq4\label{FPG}\\
&& { \rm {EV}: u^*_{R}}=\frac{\epsilon}{8},\qquad {\text at}\qquad d<4\label{FPP}.
 \end{eqnarray}
where $\epsilon=4-d$ denotes the deviation from the upper critical dimension. A final result in this approach is usually given as a series in $\epsilon$ and and an accurate quantitative result that can be compared with experimental data require at the very least the terms up to $\sim\epsilon^2$. However this is hard to achieve for a case of complex branched polymers. To overcome this problem we will use the method proposed by Douglas and Freed \cite{Douglas84}. It starts by considering a generalized scaling form  for the radius of gyration:
\begin{equation}
\langle R_g^2 \rangle = \langle R_g^2 \rangle_0 \left(\frac{2\pi N}{\Lambda}\right)^{2\nu(\eta)-1}f_p(\eta),\label{SF}
\end{equation}
where $N$ is the degree of polymerization, $\Lambda$ is the coarse-graining length scale \cite{Ohno85}, and $f_p(\eta)$ is a function that controls the solvent quality with $\eta$ being the crossover variable. It is equal 1 for $\eta=0$ (Gaussian chain) and $1+a$ for $\eta\rightarrow \infty$ which corresponds to the case of good solvent where $a$ is topology-dependent parameter. At this point it is important to note that $\left(\frac{2\pi N}{\Lambda}\right)^{2\nu(\eta)-1}$ is the same for all molecular topologies which yields that the size ratio has the following general form:
\begin{equation}
g_x = \frac{\langle R_{g,1}^2 \rangle_0}{\langle R_{g,2}^2 \rangle_0}\frac{1+a_1}{1+a_2}.
\end{equation}
This expression does not contain any approximations since the parameters $a_1$ and $a_2$ are the functions of $\epsilon$. The approximation itself comes into play when the connection between a renormalization group approach and that of the two-parameter model for a $d=3$-dimensional space is made by expression:
\begin{equation}
a=\frac{3}{4}\frac{\epsilon}{8}R_1(l,d=3)-\frac{1}{4}
\end{equation}
where $R_1(l,d=3)$ is the coefficient mentioned above but calculated for $d=3$ which corresponds to two-parameter model calculation. 

\subsection{Molecular Dynamics Simulations}
 We consider a three-dimensional, bead-spring coarse-grained model \cite{grest1987} of a dumbbell polymer which is comprised of $N$ spherical beads in each of two rings and $N_c$ beads in the linear backbone. Each bead has the size $\sigma_{LJ}$ and equal mass $m$. 
 The nonbonded interactions between monomers are taken into
account by means of the Weeks-Chandler-Anderson (WCA)
interaction, i.e., the shifted and truncated repulsive branch of
the Lennard-Jones potential given by
\begin{equation}
 V^{\rm WCA}(r) = 4\epsilon_{\rm LJ}\left[
(\sigma_{\rm LJ}/ r)^{12} - (\sigma_{\rm LJ} /r)^6 + 1/4
\right]\theta(2^{1/6}\sigma_{\rm LJ}-r).
\label{wca}
\end{equation} 
  In the above equation $r$ is denotes a distance between the centers of spherical beads, while $\epsilon$ and $\sigma_{\rm LJ}$ are chosen as units of energy and length, respectively.
  In Eq.~(\ref{wca}) is the Heaviside step function $\theta(x)=0$ or 1 for $x<0$ or $x\geq 0$.  
  The bonds between subsequent beads are described by the Kremer-Grest potential \cite{grest1986}  $V^{\rm{KG}}(r)=V^{\rm 
FENE}(r)+V^{\rm WCA}(r)$, where the first term 
represents 
by the finitely-extensible
nonlinear elastic (FENE) spring modelled by the following potential:
\begin{equation}
 V^{\rm FENE}(r)=- 0.5kr_0^2\ln{[1-(r/{r_0})^2]}.
\label{fene}
\end{equation}
with two constants $k=30\epsilon/\sigma_{\rm LJ}^2$ and $r_0=1.5\sigma_{\rm LJ}$.

The simulations where conducted using the Large-scale Atomic/Molecular Massively Parallel Simulator (LAMMPS) \cite{lammps}, that solves the Newton's equations of motion using a velocity-Verlet algorithm. The temperature $T$ was maintained by presence of the Langevin dumping term with the coefficient
$\zeta=0.5\,m\tau^{-1}$, where $\tau = \sqrt{m\sigma_{\rm LJ}^2/\epsilon}$ is the LJ time unit. The simulations were carried out in the cubic box with periodic boundary conditions in all three dimensions with equations of motion solved with the integration  step  $\Delta t = 0.005\tau$.

The initial conformations for the dumbbell polymers were generated using as a self-avoiding walk (SAW) technique generated by applying $20 (2+l) N$ symmetry operations on the simple cubic lattice \cite{Madras88} (see detail on pivot algorithm). This approach allowed to start from a more compact conformations and save computation time since the method allows for more radical changes to the trajectories per step then the MD simulation. Note that the advantage of MD simulations is better sampling of conformations when calculating size ratios \cite{Kaluzhniy22}.

The simulations where run for up to 27 molecules in the simulation box, with the steric interactions between the molecules turned off to describe a dilute solution conditions. The data for the averaging of observables was accumulated for a time period of at least three relaxation times of the corresponding systems.

We considered dumbbell polymers with degree of polymerization of rings $N=50,100,150,200,250$ and $300$ beads and degree of polymerization of linear backbones  of $N_c=N/4,N/2,N,3N/2$ and $2N$. For each of value of $N_c$ we calculated universal size ratios $g$ in the asymptotic limit, i.e. by removing the finite size effects via a least-square fitting of the form: $g_{c}(N)=g+A/N^{0.53}$ with $g$ and $A$ being fitting constants. 
In the calculation of $g$ factor (see Eq.~(\ref{gc}))  
we used the radius of gyration $R^2_{\rm linear}$ of the corresponding linear chain of the same overall degree of polymerization as dumbbell molecule. The values $R^2_{\rm linear}$ were obtained from a fitting function based on the simulations data in a range $100$ to $600$ beads where the expected scaling regime was observed. For the fitting the best known numeric values for the scaling exponent $\nu$ and the correction to scaling exponent $\Delta$ \cite{Clisby10} were taken into account. This allowed us to utilized the most accurate values of $R^2_{\rm linear}$ in the calculation of the size ratio $g$.  

\subsection{Wei's method}

\label{Wei}

A complex macromolecule can be represented as a mathematical graph (network) where monomers  are represented as vertices and chemical bonds between them as bonds of the graph. In this therms a degree of a node corresponds to the monomer functionality. 
The size and shape characteristics for any polymer network can be calculated using the Wei's method \cite{Wei}, which uses the
Kirchhoff matrix and its eigenvalues.

A polymer that consists of $N$ monomers is described by the Kirchhoff $N\times N$ matrix ${\bf K}$. All diagonal elements of this matrix $K_{ii}$ are equal to degree of vertex $i$. The non diagonal elements $K_{ij}$ are equal to either $-1$ or $0$ for $i$ and $j$ being connected or not connected, correspondingly. 
A Kirchhoff matrix of size $N\times N$ has $N-1$ non-zero eigenvalues  $\lambda_2,\ldots,\lambda_M$:
\begin{equation}
    {\bf K}{\bf Q}_i =\lambda_i {\bf Q}_i,\,\, \,\,\,\,\,i=1\ldots M 
    \end{equation}
and $\lambda_1$ is always $0$.
The size and shape characteristics within this model are given as functions of the above mentioned eigenvalues, thus the universal size ratio of  Eq.~(\ref{gc})  is defined as:
\begin{equation}
g=\frac{\sum_{j=2}^{M}1/\lambda_j^{{\rm network}}}{\sum_{j=2}^{M}1/\lambda_j^{{\rm linear}}},
\label{gwei}
\end{equation}
where $\lambda_j^{{\rm DB}}$ and $\lambda_j^{{\rm linear}}$ are the corresponding eigenvalues for the Kirchhoff matrix describing architecture of either dumbbell or a linear chain.  

The asphericity is given by the expression \cite{Wei,Ferber15}:
\begin{equation}
\langle A_d \rangle =\frac{d(d+2)}{2}\int_0^{\infty} {\rm d} y \sum_{j=2}^{M}\frac{y^3}{(\lambda_j+y^2)^2}\left[ \prod_{k=2}^{M} 
\frac{\lambda_k}{\lambda_k+y^2}\right ]^{d/2}.
\label{awei}\end{equation}

For a dumbbell architecture there are $N=(2+l)n+2$ vertices with $n$ being the number of vertices between the branching points (see Fig.~\ref{DB_sheme}) and every vertex having a degree $k > 1$. 

\section{Results and Discussion}\label{Reslts}
\subsection{The radius of gyration of a dumbbell polymer and the universal size ratio}
\begin{figure}
    \centering
    \includegraphics[width=73mm]{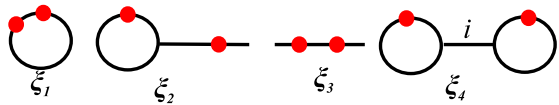}
    \caption{Schematic representation of diagrams for used in calculation of the radius of gyration in the Gaussian approximation. The polymer is depicted by solid lines and the bullets represent  the so-called restriction points $s_1$ and $s_2$}
    \label{GD}
\end{figure}    
In the continuous chain model the radius of gyration of  a dumbbell polymer is defined as:
\begin{equation}
\langle R^2_g\rangle = \frac{1}{2((2+l)L)^2}\sum_{i,j=1}^3\int_0^{L_i}\int_0^{L_j}\langle(\vec{r_i}(s_2)-\vec{r_j}(s_1))^2\rangle d s_1\,d s_2\label{RGdef}
\end{equation}
The actual calculation $R^2_g$ is performed by utilizing the following identity: 
\begin{eqnarray}
&&\langle(\vec{r}_i(s_2)-\vec{r}_j(s_1))^2\rangle = - 2 \frac{d}{d|\vec{k}|^2}\xi(\vec{k})_{\vec{k}=0},\nonumber\\
&&\xi(\vec{k})\equiv\langle{\rm e}^{-\iota\vec{k}(\vec{r}_i(s_2)-\vec{r}_j(s_1))}\rangle.\label{identity_g}
\end{eqnarray}
where the contributions to $\xi(\vec{k})$ are calculated using the diagrammatic technique. The schematic representation of the diagrams in the Gaussian approximation are displayed  in Fig.~\ref{GD}. The first two diagrams are counted twice and the later two only once. The final formula for the radius of gyration of a dumbbell polymer in the Gaussian approximation reads:
\begin{equation}
\langle R^2_g\rangle_0 =\frac{d L}{6}(l+1)(l^2+5l+3)
\end{equation}

\begin{figure}
    \centering
    \includegraphics[width=100mm]{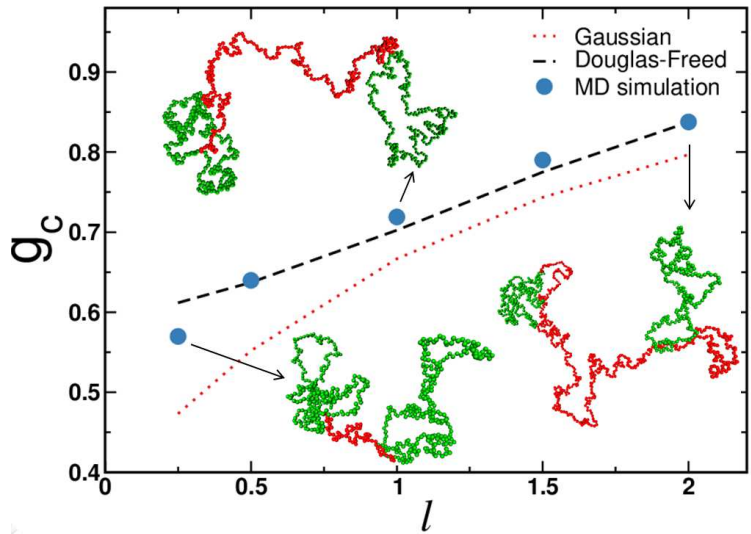}
    \caption{Relative size ratio $g_c$ of dumbbell-shaped polymers with respect to the size of the corresponding linear polymer with the same overall degree of polymerization. Data plotted   as  a function of relative degree of polymerization $l$ of linear chain monomers $N_c$ to side ring monomers $N$. The lines represent theoretical prediction obtained for the Gaussian model (dotted line) and Douglas-Freed approximation (dashed line). The circles display the results of molecular dynamics simulations. The arrows indicate simulation snapshots for dumbbell polymers with given $l$.}
    \label{gc_DB}
\end{figure} 
The corresponding universal size ratio (Eq.~(\ref{gc})) of a dumbbell polymer  in the  Gaussian approximation  is given by the expression:
\begin{equation}
g_c=\frac{(l+1)(l^2+5l+3)}{(l+2)^3}\label{g_gauss}
\end{equation}

Significantly larger set of diagrams need to be considered to estimate the radius of gyration of a dumbbell polymer with included excluded volume interactions. The calculation of $R^2_g$ in the first order of the perturbation theory, gives the following expression:

\begin{equation}
  \langle R^2_g\rangle =\frac{3 L^3(l+1)(l^2+5l+3)}{6}-u_0R_1,   
\label{Rgexplicit}
\end{equation}
where the term $R_1$ accounting for steric interactions is given by
\begin{eqnarray}
&&R_1 =
\frac{288l^2-964l-85}{72}\arcsin\left((1+4l)^{-\frac{1}{2}}\right)-\frac{\pi(192l^2-332l+47)}{72}\nonumber\\
&&-\frac{1072l^6+8308l^5+862l^4+42588l^3+36505l^2+7630l+420}{630\sqrt{l}(1+4l)(2l+1)}-\frac{\arctan((2\sqrt{l})^{-1})}{3}\nonumber\\
&&-\frac{32l^4-1088l^3-2128l^2-1684l-425}{144((2l+1)\sqrt{4l+2})}\left(\arctan\left(\frac{1+4l+\sqrt{4l+2}}{2\sqrt{l}}\right)-\arctan\left(\frac{1+4l-\sqrt{4l+2}}{2\sqrt{l}}\right)\right)
\label{R1explicit}
\end{eqnarray}

From a number of previous studies it is known that $\epsilon$-expansion provides only a qualitative agreement with  simulation and experimental data \cite{Haydukivska21,Haydukivska22,Paturej20,Douglas84}. From this reason in this work we do not provide the exact expression  for $g_c$
and limit our consideration to the Gaussian case (Eq.~(\ref{g_gauss})) and the Douglas-Freed approximation. In the latter method the data was obtained numerically using the procedure described above. 

In Fig.~\ref{gc_DB} we compare the results of our analytical calculations for the relative size $g_c$ of ideal and real dumbbell polymers with respect to size of linear polymers and compare them with the results obtained from MD simulations. The data are plotted as a function of the relative degree of polymerization $l\equiv N_c/N$, i.e. the ratio between the number of  monomers in a linear backbone to the number of monomers in a side ring. The red dotted line corresponds to the Gaussian approximation (i.e. ideal dumbbell polymer) and provides the lower boundary for $g_c$.  The black dash-line represents the analytical results of the Douglas-Freed approximation for a real dumbbell polymer, i.e. with included excluded volume interactions. 
The MD results are plotted with blue circles. We observe very good agreement between simulation data with the analytical predictions. The data obtained from all the methods consistently show that the size ratio for dumbbell architectures with $l\le 2$ is $g_c<1$ indicating a smaller molecular size of these polymers as compared to linear counterparts of the same molecular mass. The ratio $g_c$ increases with increasing $l$. For Gaussian dumbbell conformations in the limit of $l\rightarrow \infty$
it leads to the value of $g_c=1$
whereas for real dumbbell conformations the limiting value for $g_c$ obtained from the Douglas-Freed approximation is observed for $l\approx 5$ (data not shown). These  results corroborate with the recent experimental study on dumbbell-shaped polymers carried out for large  $l=8$ where it was found that $g_c\approx 1$ \cite{Doi21}  We also point out that  analytical calculations performed for  H-polymers yield the value of $g_c=1$  in the limit of $l\rightarrow \infty$, whereas for 
the case of $l = 5$ the corresponding ratio was found to be $g_c=0.95$ \cite{Pom-pom-a_22}.

\subsection{The radius of gyration radius of a dumbbell backbone and the corresponding size ratio}
Another quantity that describes equilibrium conformation of a dumbbell molecule that is  consider here is the gyration radius of dumbbell backbone defined as:
\begin{equation}
\langle r^2_g\rangle_{\rm backbone} = \frac{1}{2(l L)^2}\int_0^{L_c}\int_0^{L_c}\langle(\vec{r_0}(s_2)-\vec{r_0}(s_1))^2\rangle\label{RGbackbone}
\end{equation}

The strategy of calculation is similar to that of the full radius of gyration.   The general expression for $\langle r^2_g\rangle_{\rm backbone}$  is:
\begin{equation}
\langle r^2_g\rangle_{\rm backbone} = \frac{d l L}{6}\left(1-u_0 r_b(l,d)+\ldots\right),
\end{equation}
with $r_b(l,d)$ is the contribution  from the excluded volume interactions. The explicit formula for $\langle r^2_g\rangle_{\rm backbone}$ is provided in the Appendix, cf. Eq.~(\ref{rgb3}). 

\begin{figure}
    \centering
    \includegraphics[width=100mm]{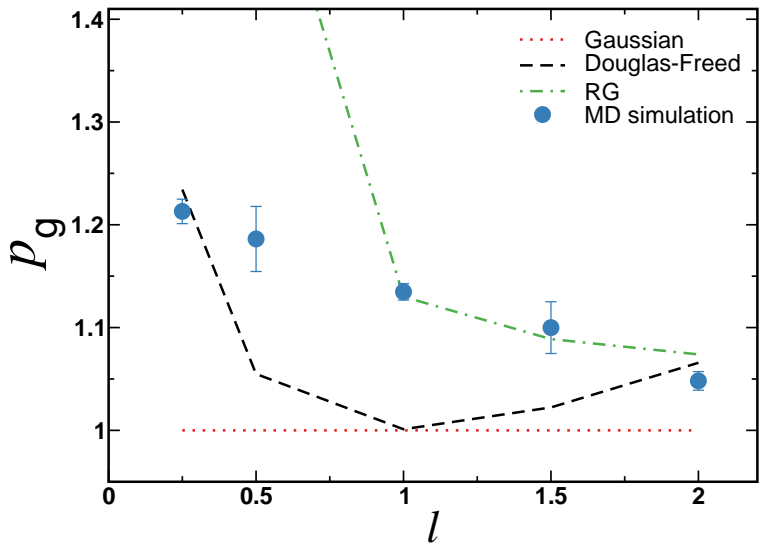}
    \caption{Relative size ratio $p_g$ of the radius of gyration of dumbbell-shaped polymers to the radius of gyration of linear polymers plotted as a function of the  relative degree of polymerization $l$. The lines represent theoretical predictions for the Gaussian conformations (dotted line),  the Douglas-Freed approximation  (dashed line) and the renormalization group calculations (dashed-dotted line). The symbols display the results of molecular dynamics simulations.}
    \label{pg_res}
\end{figure} 
To describe the influence of dumbbell side rings on the stretching of its backbone we 
introduce the relative size with respect to the corresponding size of a linear chain: 
\begin{equation}
p_g = \frac{\langle r^2_g\rangle_{\rm backbone}}{\langle R^2_g\rangle_{\rm linear}}   \label{pg} 
\end{equation}
Unlike in the case of the universal size ratio $g_c$ defined in Eq.~(\ref{gc}, the quantity $p_g$ does not depend on polymer topology for Gaussian conformations since it is simply equal to $p_g=1$.  
In our analytical calculations of dumbbell conformations with steric interactions  
we accounted for topology-dependent contributions to $p_g$  by considering the $\epsilon$-expansion and Douglas-Freed approximation. In Fig.~\ref{pg_res} we plot the values of $p_g$ as a function of $l$ obtained from theoretical methods (lines) and MD simulations (symbols). 
For $l\geq 1$, the  data calculated from $\epsilon$-expansion method (dashed-dotted line) is in good agreement with the simulation results.  However for $l\leq 1$ these results significantly overestimate MD data. The reason for this is the presence of logarithmic terms ($\propto \ln(l)$) in  the $\epsilon$-expansion. 
Note that the results obtained from Douglas-Freed approximation  (dashed line)  do not provide a good agreement with the numerical results. This indicates necessity of the second order perturbation  calculations to correctly capture stretching of the backbone.  

\subsection{Asphericity}

\begin{figure}
    \centering
    \includegraphics[width=100mm]{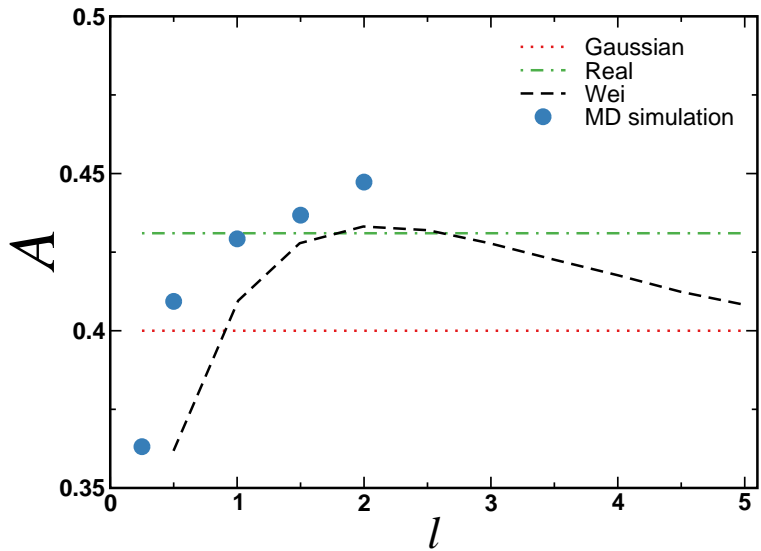}
    \caption{Asphericity $A$ of dummbell-shaped polymers plotted as a function of the relative degree of polymeryzation $l$.
    The lines represent predictions for Gaussian conformations (dotted line), real conformations  (dashed-dotted line), i.e. with included steric interactions and the calculations using Wei's methods. The symbols display the results of molecular dynamics simulations.
    }
    \label{Asph}
\end{figure}

In what follows we focus on the analysis of a shape of dumbbell polymers using asphericity factor. 
We emphasis that it is impossible to carry out path integration of the asphericity defined by Eq.~(\ref{Ad}) with the proper averaging.
From this reason we limit our consideration to the Wei's methods which provides good estimation of the asphericity for Gaussian conformations. Since in our calculations excluded volume interactions are introduce only as a small (perturbative) correction with respect to the Gaussian behavior, this approach is useful in predicting of a qualitative behavior of the molecule's shape and includes molecular architecture of polymers.  
In Fig.~\ref{Asph} we display asphericity of  dumbbell molecules 
calculated from the Wei's method and plot it as a function of $l$ (dashed line). We observe that for $l\gg 1$ asphericity decays towards the limiting value of $A=0.4$ \cite{Gaspari87} known for Gaussian linear chain (dotted line).  The similar trend is observed from our MD data (symbol) but in this case the limiting value of asphericity is $A=0.431$ (dashed-dotted line) which is known from previous studies on a linear chain in 
 good solvent \cite{Jagodzinski92}.


\section{Conclusions} \label{Con}
 In this work we have studied the conformational properties of diluted dumbbell-shaped macromolecules
 consisting of two ring polymers connected to the ends of a linear spacer.
  For this purpose we used combination of analytical calculations using field-theoretical methods and molecular dynamic simulations. 
 We have determined two relative size ratios $g_c$ and $p_g$ which are defined respectively as a ratio of the size $R_g$ of a dumbbell and the size $r_g$ of a linear spacer to the size $R_{\rm linear}$ of a linear chain of the same molecular weight.
 Our results indicate that 
 conformations of dumbbells with short linear spacers are much more compact as compared to the linear polymer coils ($g_c<0.65$).   In this structural regime side rings have the major contribution to the global dumbbell conformation and cause stretching of the spacer segment ($p_g>1$).  Increasing the degree of polymerization of a spacer 
restores its conformational flexibility ($p_g\approx 1$) and 
 gradually increases  the size of  a dumbbell molecule $(g_c\rightarrow 1)$. Finally, for spacers that are much longer than side rings the size of a dumbbell matches the size of the corresponding linear chain ($g_c=1$).  
 Our numerical data for $g_c$ are in a very good agreement with the analytical predictions.  
  The theory  correctly captures the cross-over of $g_c$ with increasing length of a spacer.  
 Our results also corroborate with the recent experiments that were carried out for dumbbells with long spacers \cite{Doi21} which reported $g_c\approx 1$.   

\begin{acknowledgments}
J.P.~and K.H.~would like to acknowledge  the support from the National Science Center, Poland (Grant No.~2018/30/E/ST3/00428) and
 the computational time at PL-Grid,  Poland.
\end{acknowledgments}

\section{Appendix}
Here we consider the details of the calculations of the contributions from the first order of perturbation theory to the partition function (Eq.~~\ref{Z}) and the corresponding calculations of the radius of gyration of a dummbbell polymer (Eq.~~\ref{R2g}). In both cases a certain set of diagrams has to be considered. For the partition function the diagrams are the same as for the Gaussian approximation of the gyration radius (see Fig.~\ref{GD}). They only difference is that we include interaction points
instead of restriction points. 
The corresponding expressions are given below:
\begin{eqnarray}
&&Z_1=u(2\pi)^{-\frac{d}{2}}(2\pi L)^{-d} L^{2-\frac{d}{2}}\frac{\Gamma\left(1-\frac{d}{2}\right)^2(2-d)}{2\Gamma(3-d)}\\
&&Z_2=u(2\pi)^{-\frac{d}{2}}(2\pi L)^{-d} L^{2-\frac{d}{2}}\left(\frac{2^{d-2}\sqrt{\pi}\Gamma(2-\frac{d}{2})}{(2-d)\Gamma(\frac{5}{2}-\frac{d}{2})}-\frac{2^{d-1}(4l+1)^{1-\frac{d}{2}}}{d-2}{\rm hypergeom}\left(\frac{1}{2}, \frac{d}{2}-1; \frac{3}{2};\frac{1}{4l+1}\right)\right)\\
&&Z_3=\frac{u(2\pi)^{-\frac{d}{2}}(2\pi L)^{-d}L_c^{2-\frac{d}{2}}}{\left(1-\frac{d}{2}\right)\left(2-\frac{d}{2}\right)}\\
&&Z_4=u(2\pi)^{-\frac{d}{2}}(2\pi L)^{-d} L^{2-\frac{d}{2}} \int_{4l+1}^{4l+2}\frac{\left(\frac{x}{4}\right)^{-\frac{d}{2}}{\rm hypergeom}\left(\frac{1}{2}, \frac{d}{2}; \frac{3}{2};\frac{1}{x}\right)}{2\sqrt{2-x+4l}}d x\\
\end{eqnarray}

Here we would like to point out the complications encounter in calculations of the term $Z_4$ which contains an integral. However since this integral does not contain divergences in respect to $\epsilon=4-d$, we can perform the expansion of the integrand. As in one loop approximation we have to consider only the terms with pole and terms $\propto\epsilon^0$. Note that we effectively calculate the expression $Z_4$ only for the $d=4$. Thus, the first order contribution into the partition function function will read:
\begin{equation}
Z_1(l,d)=2Z_1+2Z_2+Z_3+Z_4
\label{Z1appendix}
\end{equation}

\begin{figure}
    \centering
    \includegraphics[width=73mm]{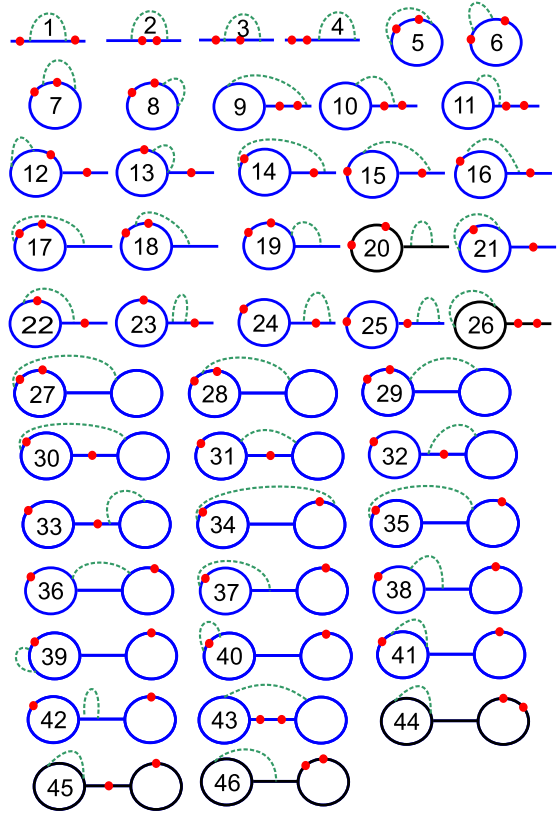}
    \caption{Schematic representation of diagrams  utilized in calculation of the radius of  gyration radius in one loop approximation. The polymer is depicted by solid lines and the bullets represent the so-called restriction points $s_1$ and $s_2$. Dash-line represents  the excluded volume interactions. }
    \label{GD1}
\end{figure} 

The calculation for the radius of gyration are conducted in the similar manner and are also complicated. Here, we have to calculate the set of diagrams displayed in Fig.~\ref{GD1}. Here, diagrams $25,\,26,\,44-46$ are reducible, as interaction points and restriction points do not share the same trajectory and thus the contributions are reduced to the product of the respective diagrams for the partition function and the Gaussian approximation for the radius of gyration. In general the calculation of the radius gyration is written as:
\begin{eqnarray}
&&\langle R^2_g\rangle = Z^{-1}(\langle R^2_g\rangle_0 - u (\textrm{Sum of diagrams}))=\langle R^2_g\rangle_0Z^{-1}\left(1-u\frac{(\textrm{Sum of diagrams})}{\langle R^2_g\rangle_0}\right)=\nonumber\\
&&\langle R^2_g\rangle_0(1+uZ_x)\left(1-u\frac{(\textrm{Sum of diagrams})}{\langle R^2_g\rangle_0}\right)=\langle R^2_g\rangle_0\left(1-u\left[\frac{(\textrm{Sum of diagrams})}{\langle R^2_g\rangle_0}-Z_x\right]\right)
\end{eqnarray}
Note that any contributions that appear in the product $Z_x\langle R^2_g\rangle_0$ cancel out. These are the diagrams mentioned above and some parts of the remaining diagrams, that in the case of the radius of  gyration are easily identifiable. As an example we consider diagrams $9-11$:
\begin{eqnarray}
&&\int^S_0\,ds\,\int^L_0\,dz\,\int^L_s\,ds_2\,\int^{s_2}_s\,ds_1\left(s_2-s_1-\frac{(s_2-s_1)^2}{s+z-\frac{s^2}{L}}\right)(s+z-\frac{s^2}{L})^{-\frac{d}{2}}+\nonumber\\
&&\int^S_0\,ds\,\int^L_0\,dz\,\int^L_s\,ds_2\,\int^{s}_0\,ds_1\left(s_2-s_1-\frac{(s-s_1)^2}{s+z-\frac{s^2}{L}}\right)(s+z-\frac{s^2}{L})^{-\frac{d}{2}}+\nonumber\\
&&\int^S_0\,ds\,\int^L_0\,dz\,\int^s_0\,ds_2\,\int^{s_2}_0\,ds_1\left(s_2-s_1\right)(s+z-\frac{s^2}{L})^{-\frac{d}{2}}
\end{eqnarray}
Since all three integrals contain the same factor $(s+z)^{-\frac{d}{2}}$ that does not depend on the restriction points this expression can be rewritten as:
\begin{eqnarray}
&&\int^S_0\,ds\,\int^L_0\,dz\,\left[\int^s_0\,ds_2\,\int^{s_2}_0\,ds_1\left(s_2-s_1-\frac{(s_2-s_1)^2}{s+z-\frac{s^2}{L}}\right)+\right.\nonumber\\
&&\left.\int^L_s\,ds_2\,\int^{s}_0\,ds_1\left(s_2-s_1-\frac{(s-s_1)^2}{s+z-\frac{s^2}{L}}\right)+\int^L_s\,ds_2\,\int^{s_2}_s\,ds_1\left(s_2-s_1\right)\right](s+z-\frac{s^2}{L})^{-\frac{d}{2}}
\end{eqnarray}
All the integrals inside the square brackets $\left[ \ldots\right]$ contain the same term under the integration, that allows to rewrite the expression as:
\begin{eqnarray}
&&\int^L_0\,ds\,\int^L_0\,dz\,\left[\int^s_0\,ds_2\,\int^{s_2}_0\,ds_1\left(s_2-s_1\right)+\int^L_s\,ds_2\,\int^{s}_0\,ds_1\left(s_2-s_1\right)+\int^L_s\,ds_2\,\int^{s_2}_s\,ds_1\left(s_2-s_1\right)\right](s+z-\frac{s^2}{L})^{-\frac{d}{2}}\nonumber\\
&&+\int^S_0\,ds\,\int^L_0\,dz\,\left[\int^s_0\,ds_2\,\int^{s_2}_0\,ds_1\left(-\frac{(s_2-s_1)^2}{s+z-\frac{s^2}{L}}\right)+\int^L_s\,ds_2\,\int^{s}_0\,ds_1\left(-\frac{(s-s_1)^2}{s+z-\frac{s^2}{L}}\right)\right](s+z-\frac{s^2}{L})^{-\frac{d}{2}}
\end{eqnarray}
The last two terms in the first line can be joined since the limits of the integration over $s_2$ are the same and the integration over $s_1$ can be presented as one integral:
\begin{eqnarray}
&&\int^S_0\,ds\,\int^L_0\,dz\,\left[\int^s_0\,ds_2\,\int^{s_2}_0\,ds_1\left(s_2-s_1\right)+\int^L_s\,ds_2\,\int^{s_2}_0\,ds_1\left(s_2-s_1\right)\right](s+z-\frac{s^2}{L})^{-\frac{d}{2}}+\nonumber\\
&&\int^S_0\,ds\,\int^L_0\,dz\,\left[\int^s_0\,ds_2\,\int^{s_2}_0\,ds_1\left(-\frac{(s_2-s_1)^2}{s+z-\frac{s^2}{L}}\right)+\int^L_s\,ds_2\,\int^{s}_0\,ds_1\left(-\frac{(s-s_1)^2}{s+z-\frac{s^2}{L}}\right)\right](s+z-\frac{s^2}{L})^{-\frac{d}{2}}
\end{eqnarray}
The similar arguments now may be presented for the case of integration over $s_2$, so the final expression reads:
\begin{eqnarray}
&&\int^S_0\,ds\,\int^L_0\,dz\,(s+z-\frac{s^2}{L})^{-\frac{d}{2}}\left[\int^L_0\,ds_2\,\int^{s_2}_0\,ds_1\left(s_2-s_1\right)\right]\nonumber\\
&&+\int^S_0\,ds\,\int^L_0\,dz\,\left[\int^s_0\,ds_2\,\int^{s_2}_0\,ds_1\left(-\frac{(s_2-s_1)^2}{s+z-\frac{s^2}{L}}\right)+\int^L_s\,ds_2\,\int^{s}_0\,ds_1\left(-\frac{(s-s_1)^2}{s+z-\frac{s^2}{L}}\right)\right](s+z-\frac{s^2}{L})^{-\frac{d}{2}}\label{L7-9}
\end{eqnarray}
Here the first term is reducible. The more interesting part is that in the first order of the perturbation theory only the diagrams that correspond to excluded volume interactions between points on the same trajectory (Diagrams $1-8,\,12,\,13,\,20,\,21,\,23-26,\,39-42,\,44,\,45$) contain poles in their $epsilon$-expansions, that are not canceled by the partition sum. These diagrams are also easy to calculate. So for the rest of the diagrams we can make the calculations for either $d=4$ or $d=3$ and by taking into account only the terms that are not canceled. This way it is easier to handle the calculations also even in this simplification an expression for the $epsilon$-expansion will still contain the integral expressions that have to be handled numerically. Results for $d=3$ are a bit more straightforward and thus are provided in the main part. As an example we present here the expressions for the radius of gyration of a dumbbell backbone for both $d=3$ and $epsilon$-expansion:
\begin{eqnarray}
&&\label{rgb3}\langle r^2_g\rangle_{backbone}(d=3) = \frac{3 l L}{6}\left(1-u_0 \left(-\frac{67}{315}-\frac{\pi(4l+1)}{24}+\frac{1}{12}(1/12)\arcsin\left((4l+1)^{-\frac{1}{2}}\right)(4l+1)\right.\right.\nonumber\\
&&\left.\left.-\frac{\sqrt{l}(176l^4-128l^3-328l^2-130l-15)}{90(1+2l)(4l+1)}-\frac{2l^4}{9}\frac{\arctan\left(\frac{4l+1+\sqrt{2+4l}}{2\sqrt{l}}\right)-\arctan\left(\frac{4l+1-\sqrt{2+4l}}{2\sqrt{l}}\right)}{(1+2l)\sqrt{2+4l}}\right)\right)\\
&&\label{rgbe}\langle r^2_g\rangle_{backbone}(d=4-\epsilon) = \frac{d l L}{6}\left(1-u_0\left(-\frac{2}{\epsilon}+\frac{13}{12}-\ln(l)+12l\int^1_0\frac{{\rm arctanh}\left((1-4z^2+4l+4z)^{-\frac{1}{2}}\right)}{(1-4z^2+4l+4z)^{\frac{5}{2}}}\,dz\right.\right.\\
&&-\frac{l\,{\rm arctanh}\left(2+4l)^{-\frac{1}{2}}\right)(8l+7)}{\sqrt{2+4l}(1+2l)}+\frac{4l\,{\rm arctanh}((4l+1)^{-\frac{1}{2}})(4l+3)}{(4l+1)^{\frac{3}{2}}}-\frac{(l-1)}{(1+2l)(4l+1)}\nonumber\\
&&-\frac{2}{5}\frac{\ln(2)(10l+3)}{l^3}-\frac{1}{5}\frac{\ln(4l+1)(40l^2-22l-3)}{l^3(4l+1)}-\frac{1}{150}\frac{1650l^3+250l^2-614l-141)}{l^3(4l+1)}\nonumber\\
&&-\frac{3}{16l^3}\int^1_0\frac{\ln\left(\frac{4l-x+1}{4l+1}\right)\left(-2(8l+3)(4l+1)^2+4(4l+1)(8l^2+12l+3)x-(112l^2+96l+18)x^2\right)}{\sqrt{x}(4l-x+1)^2}\,dx\nonumber\\
&&\left.\left.-\frac{3}{16l^3}\int^1_0\frac{\ln\left(\frac{4l-x+1}{4l+1}\right)\left((32l+12)x^3-3x^4\right)}{\sqrt{x}(4l-x+1)^2}\,dx\right)\right)\nonumber
\end{eqnarray}

\bibliographystyle{unsrt}
\bibliography{Pom-Pom}

\end{document}